# Altering the remote past

Alexander Afriat
Universiteit Utrecht
a.afriat@phys.uu.nl

ABSTRACT: An abstract treatment of Bell inequalities is proposed, in which the parameters characterizing Bell's observable can be times rather than directions. The violation of a Bell inequality might then be taken to mean that a property of a system can be changed by the *timing* of a distant measurement, which could take place in the future.

## 1 Introduction

By "remote past" I mean "the past, many miles away"; or just "long ago." Either way, the possibility of altering it is also remote, but worth considering nonetheless.

The scheme proposed here is abstract, commitment to particular semantics being deliberately avoided. Realization could be attempted with pairs of suitably correlated squids, kaons, or $B^0$-particles for instance.[1]

The observable—sometimes known as *Bell's observable*—used to violate Bell's inequality (Bell 1987) is a function of four quantities, two for one side, two for the other. These are usually angles or directions, but could also be times or quantities concerning the Hamiltonians. Provided *realism* is granted, a violation of Bell's inequality indicates the violation of an appropriate *parameter independence*; if the quantities are physical directions, a value possessed by a system can be modified by the physical rotation of a distant apparatus. If the parameters are *times*, one could conclude that the value possessed by a system at time $t$ depends on whether a measurement on a distant system is made at $t'$ or at $t''$.

## 2 Bell's inequality

Suppose an appropriate source produces many pairs $\left(\mathfrak{O}^1(k), \mathfrak{O}^2(k)\right)$ of objects, $k = 1, \ldots, N$. Each object $\mathfrak{O}^s(k)$ is assumed to possess the dichotomous property $\underline{\sigma}^s_{m,n}(k) = \pm 1$ characterized by the parameters $m$ and $n$ $(s = 1, 2)$. So for a given $s \in \{1, 2\}$ and $k \in \{1, \ldots, N\}$, the expression $\underline{\sigma}^s_{m,n}(k)$ is a function of the two arguments $m$ and $n$.

Let us define
$$\underline{B}(k) = \underline{\sigma}^1_{a,b}(k)\underline{\sigma}^2_{b,a}(k) - \underline{\sigma}^1_{a,b'}(k)\underline{\sigma}^2_{b',a}(k) + \underline{\sigma}^1_{a',b}(k)\underline{\sigma}^2_{b,a'}(k) + \underline{\sigma}^1_{a',b'}(k)\underline{\sigma}^2_{b',a'}(k),$$
whose modulus can reach 4. If we now make the assumption, which can be called 'parameter independence' that the value of $\underline{\sigma}^s_{m,n}(k)$ depends only on the first parameter

---

[1] See Afriat and Selleri (1998), Afriat (2001), Dubé and Stamp (1998a,b, 2001), Ghirardi *et al.* (1992), Selleri (1983, 1997).

$m$ and not on the second parameter $n$, we can drop the second subscripts, rewrite $\underline{B}(k)$ as

$$\underline{B}(k) = \underline{\sigma}_a^1(k)\{\underline{\sigma}_b^2(k) - \underline{\sigma}_{b'}^2(k)\} + \underline{\sigma}_{a'}^1(k)\{\underline{\sigma}_b^2(k) + \underline{\sigma}_{b'}^2(k)\}$$

and hence *halve* the bound on the modulus of $\underline{B}(k)$, from 4 to 2. The modulus of the average

$$\underline{B} = \frac{1}{N}\sum_{k=1}^N \underline{B}(k)$$

can therefore not exceed 2, which is a *Bell inequality*. We can also write

$$-2 \leq \underline{B} = \underline{P}(a,b) - \underline{P}(a,b') + \underline{P}(a',b) + \underline{P}(a',b') \leq 2,$$

where the correlation function $\underline{P}(m,n)$ is equal to

$$\frac{1}{N}\sum_{k=1}^N \underline{\sigma}_m^1(k)\underline{\sigma}_n^2(k).$$

## 3 Quantum mechanics

Unitary self-adjoint zero-trace operators on $\mathbb{C}^2$, which we shall call 'generalized Pauli operators,' are characterized by a pair of angles $(\varphi, \theta)$. For our purposes one angle is enough, so we can leave $\theta = \theta_0$ fixed and write

$$\sigma_\varphi = \sigma_{(\varphi,\theta_0)} = |\varphi_+\rangle\langle\varphi_+| - |\varphi_-\rangle\langle\varphi_-|,$$

where the $|\varphi_\pm\rangle$ are orthonormal. Turning to $\mathbb{C}^2 \otimes \mathbb{C}^2$, the value $\langle \Sigma | B | \Sigma \rangle$ involving the self-adjoint operator

$$B = \sigma_\alpha^1 \otimes \sigma_\beta^2 - \sigma_{\alpha'}^1 \otimes \sigma_\beta^2 + \sigma_\alpha^1 \otimes \sigma_{\beta'}^2 + \sigma_{\alpha'}^1 \otimes \sigma_{\beta'}^2$$

and the vector

$$|\Sigma\rangle = \frac{1}{\sqrt{2}}\left(|\varphi_+^1\varphi_-^2\rangle - |\varphi_-^1\varphi_+^2\rangle\right)$$

reaches its maximum of $2\sqrt{2}$ when the operators making up $B$ are spaced at intervals of $\pi/2$, for instance as follows:

$$\alpha = \beta - \frac{\pi}{4} = \alpha' - \frac{2\pi}{4} = \beta' - \frac{3\pi}{4}.$$

Letting $\alpha$ vanish, we can write

$$B = \sigma_0^1 \otimes \sigma_{\pi/4}^2 - \sigma_{\pi/2}^1 \otimes \sigma_{\pi/4}^2 + \sigma_0^1 \otimes \sigma_{3\pi/4}^2 + \sigma_{\pi/2}^1 \otimes \sigma_{3\pi/4}^2.$$

For any pair $\alpha, \alpha'$ of angles there is a unitary operator

$$U_{\Delta\alpha} = e^{i\Delta\alpha}|+\rangle\langle+| + |-\rangle\langle-| = e^{i\Delta\alpha|+\rangle\langle+|}$$

such that

$$\sigma_{\alpha'} = U_{\Delta\alpha}\sigma_\alpha U_{-\Delta\alpha},$$

where $\Delta\alpha$ is the difference $\alpha' - \alpha$ and the $|\pm\rangle$ are orthonormal. If the $|\pm\rangle$ happen to be the eigenvectors of a *maximal* time-independent Hamiltonian

$$H = E_+|+\rangle\langle+| + E_-|-\rangle\langle-|,$$

angles will correspond to times; then for any pair $\alpha, \alpha'$ of angles there will be a time-evolution operator

$$e^{iHt} = e^{iE_+ t}|+\rangle\langle+| + e^{iE_- t}|-\rangle\langle-|$$

such that

$$\sigma_{\alpha'} = e^{iHt}\sigma_\alpha e^{-iHt}.$$

Times can then be associated with our 'one-parameter' generalized Pauli operators (whose other parameter $\theta$ is left at some $\theta = \theta_0$); we can write

$$\sigma_{t'} = e^{iH(t'-t)}\sigma_t e^{iH(t-t')}$$

for any pair of times $t, t'$, and, returning to $\mathbb{C}^2 \otimes \mathbb{C}^2$,

$$\tilde{B} = \sigma_t^1 \otimes \sigma_u^2 - \sigma_t^1 \otimes \sigma_{u'}^2 + \sigma_{t'}^1 \otimes \sigma_u^2 + \sigma_{t'}^1 \otimes \sigma_{u'}^2,$$

where $u, u'$ are also times.

The value $\langle \Sigma | \tilde{B} | \Sigma \rangle$ can be expressed as

$$P(t,u) - P(t,u') + P(t',u) + P(t',u'),$$

where

$$P(m,n) = \langle \Sigma(m,n)|\sigma_0^1 \otimes \sigma_0^2|\Sigma(m,n)\rangle = \cos\{\Delta E(n-m)\},$$
$$|\Sigma(m,n)\rangle = (e^{iHm} \otimes e^{iHn})|\Sigma\rangle,$$

and $\Delta E$ is the difference $E_- - E_+$. At times

$$t = \frac{t_0}{\Delta E} \qquad t' = \frac{1}{\Delta E}\left(t_0 + \frac{\pi}{2}\right)$$
$$u = \frac{1}{\Delta E}\left(t_0 + \frac{\pi}{4}\right) \quad u' = \frac{1}{\Delta E}\left(t_0 + \frac{3\pi}{4}\right)$$

the value $\langle \Sigma | \tilde{B} | \Sigma \rangle$ reaches its maximum of $2\sqrt{2}$; $t_0$ is an arbitrary initial time.

If we now assume that the pairs $\left(\mathfrak{O}^1(k), \mathfrak{O}^2(k)\right)$ of objects are accurately described by $|\Sigma(m,n)\rangle$, and moreover that measurement of $\sigma_m^s$ faithfully reveals the corresponding properties $\underline{\sigma}_m^s(k)$, $k = 1, \ldots, N$, then observables represented by $B$ or $\tilde{B}$ would violate parameter independence.

## 4 Altering the remote past

Since $\underline{\sigma}_{m,n}^s(k)$ was assumed not to depend on the second subscript $n$, we dropped it and wrote $\underline{\sigma}_m^s(k)$. But now that parameter independence is once more at issue, we will sometimes need the added generality of the expression $\underline{\sigma}_{m,n}^s(k)$. Omission of the second subscript will not, however, mean that parameter independence is again assumed. Suppose the subscripts represent times. The notation $\underline{\sigma}_t^2(l) = +1$, with just the first subscript, means that the second subsystem of the $l$-th pair has the value $+1$ at time $t$; measurement of $\sigma_t^2$ would accordingly yield $+1$. The first subscript of $\underline{\sigma}_{t,m}^2(l)$, here fixed at $n = t$, represents the time pertaining to the second subsystem. The expression $\underline{\sigma}^2_{t,m}(l)$ still means, even with the second subscript, that a measurement of $\sigma^2_t$ would accurately reveal the value of $\underline{\sigma}^2_{t,m}(l)$; the second subscript indicates that the outcome of the measurement also depends on the time pertaining to the second subsystem. So much for notation.

For the modulus of

$$\underline{\tilde{B}} = \frac{1}{N}\sum_{k=1}^{N} \underline{\tilde{B}}(k)$$

to exceed 2, the modulus of

$$\underline{\tilde{B}}(k) = \underline{\sigma}^1_{t,u}(k)\underline{\sigma}^2_{u,t}(k) - \underline{\sigma}^1_{t,u'}(k)\underline{\sigma}^2_{u',t}(k) + \underline{\sigma}^1_{t',u}(k)\underline{\sigma}^2_{u,t'}(k) + \underline{\sigma}^1_{t',u'}(k)\underline{\sigma}^2_{u',t'}(k)$$

must also exceed 2 for at least one value of $k$. If we suppose that $\underline{\tilde{B}}(k) \neq 2$ for $k = k_0$, there must be at least one time $h$ such that $\underline{\sigma}^s_{h,j}(k_0) \neq \underline{\sigma}^s_{h,j'}(k_0)$, where $h, j, j' \in \{t, t', u, u'\}$. Suppose $\underline{\sigma}^1_{t',u}(k_0) \neq \underline{\sigma}^1_{t',u'}(k_0)$, and that the last two terms of $\underline{\tilde{B}}(k_0)$ are $(-1)\cdot\underline{\sigma}^2_u(k_0)$ and $(+1)\cdot\underline{\sigma}^2_{u'}(k_0)$. At first sight this seems impossible, given the meanings we have attached to the symbols involved. And perhaps it is, in which case either quantum mechanics is wrong, or expressions like $\underline{B}$ and $\underline{\tilde{B}}$ make no sense in the first place. But here we are assuming that quantum mechanics is right, and that expressions like $\underline{B}, \underline{\tilde{B}}$ do make sense, and exploring the implications. So we must wonder how it is that

$$\underline{\sigma}^1_{t',u}(k_0) = -1 \neq +1 = \underline{\sigma}^1_{t',u'}(k_0),$$

in other words that $\underline{\sigma}^1_{t'}(k_0) = -1$ when the (first) subscript of the neighbouring factor is $u$ whereas $\underline{\sigma}^1_{t'}(k_0) = +1$ when the subscript of the neighbouring factor is $u'$. Surely it makes no sense to say that $\underline{\sigma}^1_{t'}(k_0) = -1$ when $\underline{\sigma}^1_{t'}(k_0)$ is *written down* or *considered* alongside $\underline{\sigma}^2_u(k_0)$, but $\underline{\sigma}^1_{t'}(k_0) = +1$ when $\underline{\sigma}^1_{t'}(k_0)$ appears beside $\underline{\sigma}^2_{u'}(k_0)$. The dependence must have more substance to it than that, it must be more than an abstract 'notational' association. The apparatus may do no more than faithfully reveal a value that was there anyway, but surely the mere *consideration* of $\underline{\sigma}^2_{u'}(k_0)$ rather than $\underline{\sigma}^2_u(k_0)$, the fact that we express more of an interest in the former than in the latter, cannot change the value of $\underline{\sigma}^1_t(k_0)$. So *measurement* would appear to matter: if the dependence of $\underline{\sigma}^1_{t',n}(k_0)$ on the second index $n$ is to make any sense at all, the last two terms of $\underline{\tilde{B}}(k_0)$ must refer to different experimental situations. The product $\underline{\sigma}^1_{t'}(k_0)\underline{\sigma}^2_u(k_0)$ must refer to the two measurements characterized by $t'$ and $u$, the fourth term to the measurements characterized by $t'$ and $u'$. The choice of measuring $\sigma^2_{u'}$ rather than $\sigma^2_u$, and hence of revealing $\underline{\sigma}^2_{u'}(k_0)$ rather than $\underline{\sigma}^2_u(k_0)$, corresponds to a physical circumstance; the effect must be somehow due to that circumstance. Where the parameter is a direction representing the orientation of an apparatus, the circumstance is a rotation, and that's surprising enough. But now that the parameter is a time, the *very* *same* quantity is measured at times $u$ and $u'$. The apparatus remains unchanged; it does exactly the same thing, only later.

Re-writing the time-evolution operator $e^{iHt} \otimes e^{iHu}$ as $e^{i(1H)t} \otimes e^{i(g_0 H)t}$, the time on either side would be the same, the new parameter $g$, with values $g=1$ and $g=g_0$, would concern the Hamiltonians; it might represent the intensity of an appropriate field, for instance. Changing the field on one side could, if parameter independence turned out to be violated, alter the value $\pm 1$ on the other. It is surprising that physical causes as different as rotating an apparatus, waiting, changing a field can have exactly the same kind of effect. Admittedly they are all described by the same formalism; but since that formalism is about all they have in common, one might even suspect that the dependence *is* merely 'formal' or 'notational' rather than physical after all. But how can that be.

We have something of a dilemma, concerning the role of measurement. Since $\underline{\sigma}^1_{t',u}(k_0) \neq \underline{\sigma}^1_{t',u'}(k_0)$, the value $\underline{\sigma}^1_{t',n}(k_0)$ appears to depend on the second index $n$, which refers to the other object of the pair. It identifies a particular property of the other object, namely the 'time-$n$ property.' We have assumed that measurement does no more than faithfully reveal the property that was there anyway, and in no sense creates the property. But how can the time of a measurement affect a distant outcome? Where the parameter $n$ represents an angle, the effect would generally be attributed to the *physical rotation of the measuring apparatus on the other side*. But here, with times rather than angles, there seems to be no physical change worth speaking of; the experimenter just waits, and does exactly the same thing sooner rather than later. Besides, what if $\underline{\sigma}^1_{t',u}(k_0) \neq \underline{\sigma}^1_{t',u'}(k_0)$ with $t' < u, u'$? Quite apart from any *change* due to the choice of $t'$ or $t''$, does the first object have *any* value before the measurement on the other side is made? If the value $\underline{\sigma}^1_{t',n}(k_0)$ of the first object at time $t'$ does indeed depend on the time $n$ at which *measurement* is performed on the other object, what value should $\underline{\sigma}^1_{t',n}(k_0)$ be given before that second measurement? What if no measurement is made on the second object? Does the first object have any value in that case?

It is far from clear how how *waiting* can change a value possessed by an object that could be spatially and temporally remote. But let us assume it can, and consider the implications. To begin with, the properties $\underline{\sigma}^s_{m,n}(k) = \pm 1$ could be linked to larger circumstances to amplify the effects in question: By making a given measurement today at five o'clock in Utrecht, two trains passed each other without incident in Tokyo at noon on the first of January 2000. By waiting an hour and making the same measurement today at six, instead, the same trains collided at noon 1/1/2000.

So violation of parameter independence suggests that a property $\underline{\sigma}^s_t(k)$ of an object—and hence the fate of a train—can be changed by making a distant measurement at a time $t'$ rather than at another time $t''$. Nothing is said, beyond $t' \neq t''$, about the order of $t$, $t'$ and $t''$. Nature and common sense would appear to favour $t > t' > t''$ or $t > t'' > t'$ over the other four possible orderings, for it seems easier to change the future than the past. But should $t' > t > t''$, $t' > t'' > t$, $t'' > t > t'$ and $t'' > t' > t$ really be ruled out? The whole formalism, which makes no distinction between past and future, is so impartial toward all six orderings that one has to wonder. When the parameter is an angle, does it matter whether the apparatus is turned clockwise or anti-clockwise?

Where $t' > t > t''$, for instance, nothing is done at $t''$; then the value of $\underline{\sigma}^s_t(k)$ changes; then the measurement that was not made at $t''$ is made at $t'$ instead. What is it that changes $\underline{\sigma}^s_t(k)$? Is it that nothing was done at $t''$, in the past? Or is it that a measurement *will be* made at $t'$, in the future? Is it both? The distant change could depend on the difference $t' - t''$, in which case the influence would straddle the present and belong partly to the past and partly to the future.

Suppose $t'' > t' > t$. Should we just speak of a *correlation* between the outcome at $t$ and the choice of measuring at $t'$ or $t''$, or can we really identify the choice of measuring at $t'$ or $t''$ as the *cause*, the outcome at $t$ as the *effect*? Maybe the outcome at $t$, which occurs first, influences the choice. The choice of measuring at $t'$ or $t''$ could, in principle, depend on the outcome at $t$; but it can also be made independent. One can, for instance, appeal to the free will of the experimenter, who can decide whether to measure at $t'$ or $t''$ regardless of what happened at $t$; or one can rely on a random process, like a random number generator, to decide between $t'$ and $t''$. Surely a random number generator can be built whose output does not depend on the outcome of a measurement performed years before.

The effect at issue here, if indeed present, seems difficult to control, to exploit in any useful way. Suppose $\underline{\sigma}^1_t(1) = +1$. The measurement on the second particle is then made at, say, $t' > t$. We then get, say, $\underline{\sigma}^1_t(2) = -1$, and $\underline{\sigma}^1_t(3) = +1$. But what then? All we have established is that a subsequent measurement on the second particle may change $\underline{\sigma}^1_t(3) = +1$ to $\underline{\sigma}^1_t(3) = -1$. We know nothing, however, about which times will produce the change, and which will not. For instance, it could be that $\underline{\sigma}^1_{t,t'}(3) = -1$.

## 5 Relativity

An appeal to relativity hardly clarifies matters. So far we have spoken of the 'absolute' times $t$, $t'$ and $t''$. In a relativistic treatment $t = t_r(\eta^\pm)$, $t' = t_r(\eta')$ and $t'' = t_r(\eta'')$ can be considered the times of the corresponding events $\eta^\pm$, $\eta'$ and $\eta''$ with respect to some inertial system $r$, where $t = t_r(\eta^\pm)$ means that *the value $\pm 1$ is possessed at a spacetime point whose time coordinate is $t$ with respect to $r$*; and $\eta'$, $\eta''$ are the measurements made on the other object of the same pair. Although $\eta'$ and $\eta''$, which concern the same system, should not be spacelike separated, the relationship between $\eta^\pm$ and $\eta', \eta''$ is arbitrary. If $\eta', \eta''$ are in the past light cone of $\eta^\pm$, relativity would allow either one, or even both, to influence $\eta^\pm$, for the times of $\eta', \eta''$ would precede that of $\eta^\pm$ with respect to all inertial systems. But events $\eta', \eta''$ could also lie in the absolute future of $\eta^\pm$; the choice of measuring at $t'$ and not at $t''$ could then turn $\eta^\pm$ into the event $\eta^\mp$ occupying the same spacetime point in the absolute past of $\eta', \eta''$. And what if $\eta', \eta''$ are both spacelike separated from $\eta$? Then Lorentz transformations could change the order from $t_{r_1}(\eta^\pm) > t_{r_1}(\eta') > t_{r_1}(\eta'')$ to $t_{r_2}(\eta') > t_{r_2}(\eta^\pm) > t_{r_2}(\eta'')$ to $t_{r_3}(\eta') > t_{r_3}(\eta'') > t_{r_3}(\eta^\pm)$, or from $t_{r'_1}(\eta^\pm) > t_{r'_1}(\eta'') > t_{r'_1}(\eta')$ to $t_{r'_2}(\eta'') > t_{r'_2}(\eta^\pm) > t_{r'_2}(\eta')$ to $t_{r'_3}(\eta'') > t_{r'_3}(\eta') > t_{r'_3}(\eta^\pm)$. So can an influence going from the past to the future be turned into an influence going from the future to the past, or from the past and future to the present, by a Lorentz transformation? Maybe it makes no sense to speak of *past* and *future* in cases disallowed by relativity theory in the first place.

## 6 Conclusions

The emphasis has been not just on altering the past, but on doing so by *waiting*. Of course the possibility of changing the past is not to be taken too seriously; here it is only viewed as a consequence of quantum mechanics together with an appropriate kind of realism. The whole matter can be treated as a *reductio* argument against realism or quantum mechanics; or, if all the assumptions are to be taken seriously, I suppose one could actually wonder about altering the past, and start thinking about realizations. The possibility is of course highly paradoxical, and difficult to reconcile with most received ideas about causality.

I wish to thank Dennis Dieks, Richard Gill, Janneke van Lith, Fred Muller, Philip Stamp and Jos Uffink for many fruitful discussions.